# Stability and superconductivity of 4*d* and 5*d* transition metal high-entropy alloys

Alexander J. Browne,[1,*] Denver P. Strong[1] and Robert. J. Cava[1]

[1]Department of Chemistry, Princeton University, Princeton, New Jersey 08544, USA.

[*]Present address: School of Chemistry, University of St Andrews, North Haugh, St Andrews, KY16 9ST, UK. Email: ajb58@st-andrews.ac.uk.

## Abstract

We report the synthesis of new equimolar high-entropy alloys (HEAs) formed from five or six 4*d*/5*d* transition metals that are each from a different Group of the Periodic Table. These include MoReRuRhPt and MoReRuIrPt, which have a valence electron count (VEC) of 8 and crystallise in the hexagonal close-packed (hcp) structure, making them isoelectronic and isostructural to the elements Ru and Os and the high-pressure phase of the well-known HEA CrMnFeCoNi. Analogous HEAs with VECs of 7, 6 and 5 crystallise with multiple phases, which we rationalise using the atomic size difference, $\delta$, and the electronegativity difference, $\Delta\chi$, of the constituent elements. Finally, we find our hcp and mixed hcp-$\sigma$ phase HEAs to be superconductors, and compare their critical temperatures to those of isostructural elements and other HEAs that have recently been reported.

## Introduction

Since they were first reported in 2004, high-entropy alloys (HEAs) have attracted ever-increasing attention as prospective next-generation materials.[1–3] Unlike conventional alloys, which typically consist of a single principal element doped with small amounts of others to enhance a desirable property, HEAs are mixtures of multiple principal elements – generally five or more – in approximately equal proportions, which are stabilised as a solid solution on a simple lattice by the high configurational entropy. This unconventional approach to alloy design allows for the synthesis of new materials whilst also providing an extremely large compositional space within which their properties can be tuned. Significantly, those properties can match – and in fact often exceed – those of conventional alloys.[4] Whilst initially focused on developing their exceptional mechanical capabilities, current work is also motivated by the discovery of HEAs that exhibit functional properties such as soft magnetism,[5] magnetocaloric effects,[6] thermoelectric behaviour,[7] and superconductivity.[8]

Superconductivity in a HEA was first observed in $Ta_{0.34}Nb_{0.33}Hf_{0.08}Zr_{0.14}Ti_{0.11}$ in 2014.[9] Its behaviour illustrates how the high-entropy approach can be used to design materials with enhanced properties, because its superconducting critical temperature ($T_c$), 7.27 K, exceeds the weighted average of the critical temperatures of the constituent elements by more than 50%. $Ta_{0.34}Nb_{0.33}Hf_{0.08}Zr_{0.14}Ti_{0.11}$ and other early HEA superconductors have the simple body-centred cubic crystal structure;[10,11] however, the variety of known structure types is growing increasingly diverse (Figure 1). Known examples now include HEAs with cubic B2 (CsCl), α-Mn -type and A15 structures,[12–16] close-packed and Laves-superstructure hexagonal structures,[17–20] and tetragonal $\sigma$ and $CuAl_2$-type structures.[21–24]

The phase, or mixture of phases, formed by a multicomponent system at thermodynamic equilibrium is that which minimises the Gibbs free energy, $\Delta G$:

$$\Delta G = \Delta H - T\Delta S \ (1)$$

If that phase is an ideal solid solution, the entropic contribution to $\Delta G$ is the mixing entropy, $\Delta S_{mix}$:

$$\Delta S_{mix} = -R \sum_i x_i \ln x_i \ (2)$$

where $x_i$ is the mole fraction of element $i$ and $R$ is the gas constant. Furthermore, in the ideal case the corresponding enthalpic contribution $\Delta H_{mix}$ is negligible; as such, the formation of a HEA is driven by the maximisation of $\Delta S_{mix}$, which is achieved by randomly distributing many different elements over the single crystallographic site of a simple crystal structure. However, simply mixing a large number of elements together does not guarantee that a high-entropy phase will form, as in a real mixture there are likely to be pairs of elements for which $\Delta H_{mix}$ is non-negligible.[25] Instead, a HEA will only form if its entropically-dominated $\Delta G$ is not outweighed by that of another product (or combination of products) for which $\Delta G$ has a large enthalpic driving force.[26] In principle, therefore, predicting whether a combination of elements will successfully form a HEA requires $\Delta G$ to be calculated for all possible products of that combination, but to do so for a multicomponent mixture is non-trivial. Instead, various empirical proxies have been put forward as guides to help predict whether a particular mixture will form a HEA. These proxies include the atomic size difference, $\delta$:

$$\delta = 100 \times \sqrt{\sum_i x_i \left(1 - \frac{r_i}{\sum_i x_i r_i}\right)^2} \ (3)$$

where $r_i$ is the atomic radius of element $i$,[27] and the electronegativity difference, $\Delta\chi$:

$$\Delta\chi = \sqrt{\sum_i x_i \left(\chi_i - \sum_i x_i\chi_i\right)^2} \quad (4)$$

where $\chi_i$ is the Pauling electronegativity of element $i$.[28] Single-phase HEAs are most likely to form when the value of $\delta$ is small, and these are likely to crystallise with a simple crystal structure such as body-centred cubic (bcc), face-centred cubic (fcc) or hexagonal close-packed (hcp) when $\Delta\chi$ is also small. However, single-phase HEAs can still be realised when $\Delta\chi$ is large, though they are likely to have intermetallic structures in which there are multiple crystallographic sites.[29] The different sites can then be preferentially occupied by the species with very different electronegativities such that the prevalence of non-zero $\Delta H_{mix}$ interactions is minimised.

Where a HEA phase does form, the structure it adopts is also influenced by the valence electron count (VEC) – that is, the average number of valence electrons per atom in the composition. Of the three simple crystal structures, single-phase hcp HEAs typically have a VEC of 3 or less, bcc ones have a VEC between 4 and 6, and fcc ones a VEC greater than 8.[30] However, a small number of hcp HEAs have also been found in a window of higher VEC, coincident with the elements Tc, Re, Ru and Os in Groups 7 and 8 of the Periodic Table that themselves crystallise with the hcp structure.[31,32] Furthermore, in addition to influencing the structure of alloys, VEC also influences their tendency to exhibit superconductivity. As initially observed by Matthias,[33] the $T_c$ of transition metals and their alloys varies with VEC and shows two maxima, at VECs of around 4.5 and 6.5. $T_c$ in different families of superconducting HEAs also shows a dependence on VEC,[10] but those studied more extensively so far have VECs closer to the first maximum (for example, the VEC of $Ta_{0.34}Nb_{0.33}Hf_{0.08}Zr_{0.14}Ti_{0.11}$ is 4.67).[34] We have therefore targeted the synthesis of new high-VEC hcp HEAs, so that the structures and properties of HEAs in this regime can be better understood.

## Experimental

All samples were prepared by arc-melting of the elements (all supplied by Alfa Aesar, with purities of 99.9% or greater). In most cases, the material to be melted was prepared by grinding together stoichiometric quantities of the powders of the appropriate elements in an agate pestle and mortar and pressing this mixture into a pellet. However, this approach was modified in two cases. Firstly, where Pd was used it was provided with a 5% molar excess to compensate for its relatively low boiling point; and secondly, where Zr or Au were used the sample was prepared by melting together small pieces of all the constituent elements. In all cases, samples of approximately 500 mg were melted under argon on a water-cooled copper hearth after a zirconium getter had been heated to clean the atmosphere of residual oxygen and moisture. Each sample was turned over and melted again at least two times to aid the mixing of the constituents. The mass of each sample lost during the arc-melting was 2% or less of the stoichiometric amount expected with the exception of 5.2% for the Au-containing sample, for which unincorporated Au was found to evaporate and recondense on the copper hearth.

After arc-melting the samples were broken into small pieces using steel bolt-cutters, and the smaller pieces were ground to a powder with a pestle and mortar. Powder X-ray diffraction data were collected at room temperature using a Bruker D8 Advance ECO diffractometer operating in the Bragg-Brentano geometry with Cu Kα radiation. Profile fitting of the powder diffraction patterns to extract lattice parameters was done by the Le Bail method using GSAS.[35] Larger sample pieces were shaped into bars 2-4 mm in length with a diamond file, and these were used for measurements of electrical resistance. These measurements were done with a Quantum Design Dynacool PPMS, using a standard four-probe technique with Pt wires and Ag-paste contacts on the sample. An adiabatic demagnetisation insert was used to reach temperatures below 1.8 K.

## Results & discussion

### *Phase characterisation*

Structural analysis of all the HEAs we report is detailed in the following three subsections and summarised in Table 1, where values of $\Delta S_{mix}$, $\delta$ and $\Delta\chi$ calculated using Equations 2, 3 and 4 are provided for each. Unless otherwise stated, specified alloy compositions are equimolar: for example, NbMoReRuRh = $Nb_{0.2}Mo_{0.2}Re_{0.2}Ru_{0.2}Rh_{0.2}$. However, a composition written as e.g. Nb-Mo-Re-Ru-Rh refers to a family of materials composed of those elements in varying amounts.

**VEC = 8.** Four HEAs – MoReRuRhPd, MoReRuRhPt, MoReRuIrPd and MoReRuIrPt – were prepared by mixing one 4*d*/5*d* element from each of Groups 6 to 10 in equimolar ratios. These combinations give a VEC of 8, making them isoelectronic with the Group 8 elements Ru and Os. These elements crystallise with the hcp structure, and (perhaps surprisingly) so to do binary phases of Mo (bcc) with Rh, Ir, Pd and Pt (all fcc) in compositional ranges that give a VEC of around 8.[36] It was therefore anticipated that these four HEAs would also crystallise with the hcp structure, with the five elements statistically occupying the single crystallographic site of that structure.

Indexing the powder X-ray diffraction patterns confirms that MoReRuRhPt and MoReRuIrPt do indeed crystallise as a single hcp phase (Figure 2(a)-(b)). These are the first known HEAs for these combinations of elements, and join the very few known examples of equimolar transition metal HEAs that crystallise with the hcp structure.[31] They are, in other words, high-entropy analogues of the Group 8 elements Ru and Os, despite containing only 20 at.% of one these. Furthermore, as will be discussed later, these HEAs are superconductors with critical temperatures consistent with those of Ru and Os.

An interesting comparison of these HEAs can also be drawn with their 3*d* analogue, CrMnFeCoNi, which crystallises from the melt in the fcc structure but converts to an hcp polymorph at high pressures.[37,38] This mirrors the behaviour of the 3*d* Group 8 element, Fe, which has a high-pressure hcp allotrope $\varepsilon$-Fe. Ab-initio calculations have suggested that the hcp polymorph of CrMnFeCoNi is actually the thermodynamically-stable form of this material but that the magnetic contribution to the total $\Delta S$, which increases with temperature, stabilises the fcc phase at higher temperatures.[39] As such, the fcc phase is the one that initially crystallises from the melt, then the rapid cooling of the melt that it typical for arc-melting and other methods commonly used to synthesise HEAs results in it being kinetically trapped.[38] However, compressing the material broadens the valence *d*-band and suppresses the local magnetic moments, hence the hcp phase is stabilised at high pressures.[37] 4*d* and 5*d* transition metals have intrinsically broader valence bands than their 3*d* counterparts, and in the elements in our HEAs this must be to sufficient degree that an fcc polymorph of these HEAs is not energetically competitive even at ambient pressure, hence they crystallise directly as hcp.

The two Pd-containing HEAs, MoReRuRhPd and MoReRuIrPd, also each crystallise with an hcp phase (Figure 2(c)-(d)). However, whilst their Pt-containing analogues are equimolar single phases, these materials retain a Pd impurity. Although a Pd excess was used in their preparation the mass lost during arc-melting was large enough that the residual Pd cannot be a remnant of that excess. Instead, the presence of a Pd impurity can be understood from the perspective of Pd being less soluble than Pt in a mixture of the remaining elements. The synthesis of the HEAs discussed here is effectively the dissolution of Pd or Pt into ‘MoReRuRh’ or ‘MoReRuIr’ – systems that would have a VEC halfway between those of Ru and Re, and likely also the same hcp structure. Binary phase diagrams reveal that Pd is much less soluble in than Pt in Ru – Ru can dissolve no more than 8 at.% of Pd but up to 22 at.% of Pt.[36] For Re the difference is even greater – 6 at.% Pd to 44 at.% Pt.[36] By extension, these differences are likely reasonably indicative of the relative solubilities of Pd and Pt in the HEA mixtures. Pt can evidently dissolve sufficiently to give equimolar – and hence single-phase – HEAs, $[MoReRu(Rh/Ir)]_{0.8}Pt_{0.2}$, but Pd cannot, resulting in hcp HEA phases with non-equimolar compositions, $[MoReRu(Rh/Ir)]_{0.8+x}Pd_{0.2-x}$, and residual unincorporated Pd.

**VEC = 7, 7.5.** Following the successful preparation of VEC = 8 HEAs with equimolar combinations of 4*d*/5*d* transition metals from Groups 6 to 10, the synthesis of VEC = 7 HEAs with equimolar combinations of 4*d*/5*d* transition metals from Groups 5 to 9 was also explored. Like their Group 8 counterparts, the 4*d* (Tc) and 5*d* (Re) Group 7 elements have the hcp structure.

Two VEC = 7 HEAs, NbMoReRuRh and NbMoReRuIr, were prepared. X-ray diffraction reveals that in both cases the expected hcp phase is present; however, it coexists with a $\sigma$ phase (Figure 3(a)-(b)). $\sigma$ phases, which have a large tetragonal unit cell containing 30 atoms distributed over five crystallographic sites (Figure 1(c)), are common binary intermetallics of transition metals and have been identified in numerous HEAs as well.[40] A review of predominantly-3*d* transition metal HEAs has suggested that they are susceptible to forming a $\sigma$ phase when the VEC is between 6.88 and 7.86,[41] and VCrMnFeCo – the 3*d* analogue of the VEC = 7 HEAs discussed here – crystallises as a single $\sigma$ phase.[42] A more recent study of Ta-Mo-W-Re-Ru HEAs has suggested that a large $\Delta\chi$ is also important if a HEA is to adopt the multi-site $\sigma$ structure instead of a simple single-site one, indicating that it should fall between 0.19 and 0.24.[21] The $\Delta\chi$ values of NbMoReRuRh and NbMoReRuIr are, respectively, 0.249 and 0.234 – close enough to the suggested range that the formation of a $\sigma$ phase is likely, though the Ta-Mo-W-Re-Ru HEAs used to determine the $\Delta\chi$ criterion all crystallise as a single $\sigma$ phase. The formation of mixed hcp and $\sigma$ phases by NbMoReRuRh and NbMoReRuIr may therefore be a result of their higher values of $\delta$ – 2.56 and 2.46 respectively, compared to 1.75-2.37 for the Ta-Mo-W-Re-Ru HEAs – in addition to their large $\Delta\chi$, because a multi-phase mixture provides an even greater diversity of crystallographic sites for the atomic pairs with non-negligible $\Delta H_{mix}$ to be distributed over. However, the infeasibility of determining the occupancies of multiple sites by multiple elements that have similar X-ray scattering lengths by X-ray diffraction means that the precise chemical compositions of the two phases in these HEAs, and their relative fractions, have not yet been determined.

HEAs are by no means limited to compositions that give an integer VEC, and the six-element equimolar HEAs NbMoReRuRhPt and NbMoReRuIrPt, which have a VEC of 7.5, were also prepared. Increasing the number of equimolar components from five to six increases $\Delta S_{mix}$ by 11% and this additional entropic stabilisation appears to be influential in these systems, because although their values of $\delta$ (2.42 and 2.32) and $\Delta\chi$ (0.246 and 0.236) remain large there is no $\sigma$ phase. However, profile-fitting of the powder diffraction patterns does reveal a subtle separation of two hcp phases (Figure 3(c)-(d)). As illustrated with the earlier discussion of the CrMnFeCoNi system, the phase obtained from arc-melting followed by rapid cooling is not necessarily the one that is thermodynamically stable at room temperature; instead, it is likely to be the one that was thermodynamically stable upon initial solidification, which is then kinetically trapped by that rapid cooling. At higher temperatures the $T\Delta S$ contribution to $\Delta G$ is increasingly substantial (Equation 1), hence the stable phase at the melting point is likely to be the one that has the largest $\Delta S$ – in other words, the high-entropy solid solution. In comparison to NbMoReRuRh and NbMoReRuIr the addition of a sixth element to NbMoReRuRhPt and NbMoReRuIrPt increases $\Delta S_{mix}$; however, as the contribution of this entropic bonus to $\Delta G$ is temperature dependent it becomes less as the material cools, meaning that a faster cooling rate is required if the HEA phase is to be trapped successfully. The absence of a $\sigma$ phase in NbMoReRuRhPt and NbMoReRuIrPt shows that increasing $\Delta S_{mix}$ can overcome the enthalpic effects encoded in their large values of $\delta$ and $\Delta\chi$; however, the observed phase separation indicates that the cooling rate provided by the conventional water-cooled copper hearth we used was insufficient for successful kinetic trapping. Employing a synthesis method with a faster cooling rate, such as splat-quenching, should allow these six-element materials to be obtained as true single-phase HEAs.

**VEC = 5, 6.** The materials described so far show that preparing HEAs from combinations of elements that are all from different Groups of the Periodic Table works well so long as the chosen combinations of elements give values of $\delta$ and $\Delta\chi$ that can be balanced by $\Delta S_{mix}$. The restrictions imposed on HEA formation when the values of these parameters become too large become more evident for lower-VEC combinations. Whereas the 4*d* and 5*d* elements of Groups 7 and 8 crystallise as hcp, those of Groups 5 and 6 – Nb, Ta, Mo and W – are bcc. ZrNbMoReRu – a VEC = 6 combination of elements from Groups 4 to 8 – has a major bcc phase. However, it is accompanied by B2 (CsCl) and C14 (hexagonal Laves) phases (Figure 4(a)). A B2 phase exists in the equimolar regions of the Zr-Ru and Nb-Ru phase diagrams, whilst binary $ZrRe_2$ is a C14 phase.[36] As detailed in Table 1, the values of $\delta$ and $\Delta\chi$ for ZrNbMoReRu are much larger than those of the single-phase HEAs described above, suggesting a sufficient prevalence of atomic pairs with non-zero $\Delta H_{mix}$ that there is no way for even a large $\Delta S_{mix}$ to overcome them and form a single HEA phase. YZrNbMoRe – in which elements from Groups 3 to 7 are combined to give a VEC of 5 – has even larger values of $\delta$ and $\Delta\chi$, and though a bcc phase is formed in this material there are also C15 (cubic Laves) and $\sigma$ phases, plus elemental Y (hcp) and Mo (a second bcc phase) that are not incorporated (Figure 4(b)). C15 $ZrMo_2$, and Nb-Re and Mo-Re $\sigma$ phases, are found in the binary phase diagrams; furthermore, similar to Pd in the previous discussion of the VEC = 8 HEAs, Y is poorly miscible with the other elements in this HEA composition.[36]

It should be noted that the limitations of too-large values of $\delta$ and $\Delta\chi$ also still apply for HEAs with VECs higher than 5 and 6. Equimolar NbMoRuPtAu has the same VEC and same $\Delta S_{mix}$ as MoReRuRhPt and MoReRuIrPt, but unlike these systems does not crystallise as an hcp solid solution. Instead, a mixture of $\sigma$ and orthorhombic $MoPt_2$-type phases is formed along with unincorporated Au (Figure 4(c)). The values of $\delta$ and $\Delta\chi$ for NbMoRuPtAu are large, and the binary phase diagrams show that Au has negligible miscibility with Mo and Ru.[36] Therefore, whilst our new HEAs MoReRuRhPt and MoReRuIrPt show that high-entropy analogues of some elements can be prepared without those elements being a large component of the alloy, design consideration is required to find the subset of HEA compositions that can actually realise this.

*Superconducting properties*

Finally, electrical resistance measurements allow the hcp and mixed hcp-$\sigma$ (VEC = 8, 7.5 and 7) HEAs described above to be identified as superconductors. The first superconducting HEAs adopting these structure types were recently reported: hcp phases in non-equimolar Mo-Re-Ru-Rh-Ti, Nb-Mo-Ru-Rh-Pd and Nb-Mo-Re-Ru-Rh systems,[17,19,20] and $\sigma$ phases in Ta-Mo-W-Re-Ru and Ta-Mo-Cr-Re-Ru ones.[21,22] The VECs of the synthesised materials in these families range from 6.6 to 7.5, and the reported values of their superconducting critical temperatures $T_c$ range from 2.1 to 9.1 K. Our HEAs become superconducting at similar temperatures (Figure 5). All eight materials have a bulk superconducting ground state below $T_c$ in which the electrical resistance $R$ is zero. Taking $R_{8K}$ as representative of the low-temperature resistance of the normal metallic state of each material (the resistivities of the samples at 8 K are on the order of $10^{-5}$ Ω.cm), $T_c$ was taken to be the temperature at which $R/R_{8K} = 0.5$ and ranges from 0.82 to 6.35 K. Slight discontinuities in some of the measured transitions indicate microstructural inhomogeneity in the relatively large pieces of sample used for the measurements.

As in superconducting elements, the superconductivity observed in HEAs so far generally follows the BCS mechanism,[8,9] for which $T_c$ is dependent on electron-phonon interactions and the density of states at the Fermi level, $N(E_F)$. Considering isoelectronic systems first, the critical temperatures of our VEC = 8 HEAs can be compared to those of the Group 8 elements

Ru and Os (Figure 6(a)). $T_c$ shows a positive dependence on the volume of the hcp unit cell, as this increases $N(E_F)$. That said, for HEAs the variation of $T_c$ in this rigid-band model for $T_c$ should be supplemented with a consideration of the effects of both disorder, which places the phenomenology of superconductivity in HEAs between that of truly crystalline and fully amorphous materials,[10] and composition, which determines how the different interactions between different elements that favour or disfavour superconductivity are balanced.[43] Furthermore, only the single-phase HEAs MoReRuRhPt and MoReRuIrPt are strictly isoelectronic with Ru and Os. As discussed above, the hcp phases in MoReRuRhPd and MoReRuIrPd are actually Pd-deficient $[MoReRu(Rh/Ir)]_{0.8+x}Pd_{0.2-x}$, and as such their VECs must really be less than 8. $T_c$ has a dependence on VEC because $N(E_F)$ is dependent on the filling of the valence band, and as Matthias observed, the effect of this in a given family of materials is that $T_c$ typically reaches a maximum when the VEC is around 6.5 and then decrease for VECs larger than that.[33] This can be seen in the $T_c$ of Group 7 Re (1.695 K) being more than twice that of Group 8 Ru (0.478 K) or Os (0.62 K).[44] A similar correlation with their reduced VECs would explain why the $[MoReRu(Rh/Ir)]_{0.8+x}Pd_{0.2-x}$ HEAs have critical temperatures more than 50% higher than those of their Pt-containing analogues. As Pd is not a bulk superconductor the residual Pd in these materials does not influence $T_c$.

The critical temperatures of our HEAs, and their dependence on VEC, can also be compared to those of the previously reported superconducting hcp and $\sigma$ HEAs (Figure 6(b)). A Matthias-like trend can be seen in our materials and others, with $T_c$ decreasing as the VEC is increased from 7 to 8. A dependence of $T_c$ on composition can also be seen. Of particular relevance for comparison to our materials is the non-equimolar $Nb_{0.05}Mo_{0.35-x}Re_{0.15+x}Ru_{0.35}Rh_{0.10}$ family of hcp HEAs.[20] These have VECs ranging from 7.10 to 7.25, and if a VEC = 7 member could be prepared it seems likely that it would have a significantly higher $T_c$ than our equimolar VEC = 7 HEAs (and may also crystallise as a single hcp phase). In fact, this comparison of compositionally-related HEAs illustrates one nuance of HEA design. Their multicomponent nature allows a wide scope for the tailoring of composition-property relationships and though equimolar systems maximise $\Delta S_{mix}$ they need only be the starting point of explorations for new materials with exciting functional properties.[4]

## Conclusions

We have described the single- and multi-phase HEAs formed by combinations of 4*d* and 5*d* transition metals taken from different Groups of the Periodic Table. The hcp structure, often only considered stable for HEAs when the VEC less than 3, can also be adopted for high VECs around 7-8, and our approach to HEA design has allowed us to add to the small number of known high-VEC hcp HEAs. However, limits on this approach are imposed by the large differences of atomic size and electronegativity that can exist between transition metals from different Groups. When these differences are too large intermetallic phases become energetically competitive with high-entropy solid solutions and multi-phase mixtures are formed.

The hcp and mixed hcp-$\sigma$ HEAs we have synthesised exhibit superconductivity with critical temperatures of up to 6.35 K. As the number of high-VEC superconducting HEAs increases, trends such as the Matthias-like dependence of $T_c$ on VEC are being revealed. A family of hcp HEAs that remains stable over a wide range of VECs would allow a more detailed understanding to be developed, as has been done for bcc HEAs of the Group 4 and 5 transition metals.[10]

Both the structures and superconducting states of those bcc HEAs also show remarkable resilience to applied pressure.[45] The hcp HEA $Ir_{0.19}Os_{0.22}Re_{0.21}Rh_{0.20}Ru_{0.19}$ remains mechanically stable to at least 45 GPa,[46] and the hexagonal polymorph of the 3*d* VEC = 8 HEA CrMnFeCoNi is stabilised at high pressures,[37,38] but it is not yet known whether the superconducting states of hcp HEAs show the same robustness to pressure as those in bcc systems can.[47] The high-pressure polymorphism of CrMnFeCoNi perhaps has further relevance too, because if the reverse transition, from hcp to fcc, could be induced in our VEC = 8 hcp HEAs – perhaps by chemical substitution rather than the application of pressure – it may provide a route to the discovery of superconducting HEAs with the fcc structure that have so far proven to be elusive.[34]

Finally, the synthesis of new hcp HEAs has relevance beyond studies of superconductivity. Because there are currently fewer examples the mechanical properties of hcp HEAs have been less well-studied than those of their bcc and fcc counterparts, though the high-pressure stability of $Ir_{0.19}Os_{0.22}Re_{0.21}Rh_{0.20}Ru_{0.19}$ suggests that they have mechanical properties just as exceptional as those of other HEAs.[3] Furthermore, hcp HEAs with compositions similar to those we have deliberately synthesised are also formed in the waste products of nuclear fission reactors.[32] Developing an understanding of the chemical and physical properties of these materials is therefore of multi-disciplinary interest.

## Conflicts of interest

There are no conflicts to declare.

## Acknowledgements

This work was supported by the Gordon and Betty Moore Foundation, EPiQS initiative, grant GBMF-9006.

Table 1: Summary of all the HEAs reported here. Values of $\Delta S_{mix}$, $\delta$ and $\Delta\chi$ were calculated according to Equations 2, 3 and 4, with values for $r_i$ taken from [25] and values for $\chi_i$ being the standard Pauling electronegativities. The lattice parameters of the identified phases are taken from the Le Bail fits in Figures 2, 3 and 4.

| **Composition** | **VEC** | $\boldsymbol{\Delta S_{mix}}$ | $\boldsymbol{\delta}$ | $\boldsymbol{\Delta\chi}$ | **Phases** | **Lattice parameters (Å)** |
|---|---|---|---|---|---|---|
| MoReRuRhPt | 8 | 1.61$R$ | 1.57 | 0.140 | hcp | a = 2.75040(6), c = 4.39489(9) |
| MoReRuIrPt | 8 | 1.61$R$ | 1.47 | 0.130 | hcp | a = 2.75539(7), c = 4.39796(10) |
| MoReRuRhPd | 8 | 1.61$R$ | 1.20 | 0.130 | hcp<br>Pd (fcc) | a = 2.74580(9), c = 4.38960(13)<br>a = 3.87041(31) |
| MoReRuIrPd | 8 | 1.61$R$ | 1.10 | 0.117 | hcp<br>Pd (fcc) | a = 2.75113(8), c = 4.38682(12)<br>a = 3.88427(23) |
| NbMoRuPtAu | 8 | 1.61$R$ | 2.98 | 0.308 | $MoPt_2$-type<br>$\sigma$<br>Au (fcc) | a = 2.7734(4), b = 8.3789(12), c = 3.9685(5)<br>a = 9.6411(12), c = 4.9419(7)<br>a = 4.0801(3) |
| NbMoReRuRhPt | 7.5 | 1.79$R$ | 2.42 | 0.246 | hcp-I<br>hcp-II | a = 2.7634(1), c = 4.4448(2)<br>a = 2.7788(1), c = 4.4791(2) |
| NbMoReRuIrPt | 7.5 | 1.79$R$ | 2.32 | 0.236 | hcp-I<br>hcp-II | a = 2.7618(2), c = 4.4360(3)<br>a = 2.7777(2), c = 4.4757(3) |
| NbMoReRuRh | 7 | 1.61*R* | 2.56 | 0.249 | hcp<br>*σ* | a = 2.7728(3), c = 4.4766(5)<br>a = 9.6261(18), c = 4.9752(13) |
| NbMoReRuIr | 7 | 1.61*R* | 2.46 | 0.234 | hcp<br>*σ* | a = 2.7755(2), c = 4.4739(3)<br>a = 9.6524(10), c = 4.9781(6) |
| ZrNbMoReRu | 6 | 1.61*R* | 7.33 | 0.333 | bcc<br>B2<br>C14 | a = 3.17887(8)<br>a = 3.21666(9)<br>a = 5.25453(15), c = 8.63264(29) |
| YZrNbMoRe | 5 | 1.61*R* | 11.63 | 0.350 | C15<br>bcc<br>Mo (bcc)<br>Y (hcp)<br>*σ* | a = 7.6382(4)<br>a = 3.2249(2)<br>a = 3.1481(2)<br>a = 3.6442(5), c = 5.7210(9)<br>a = 9.4637(14), c = 4.9129(8) |

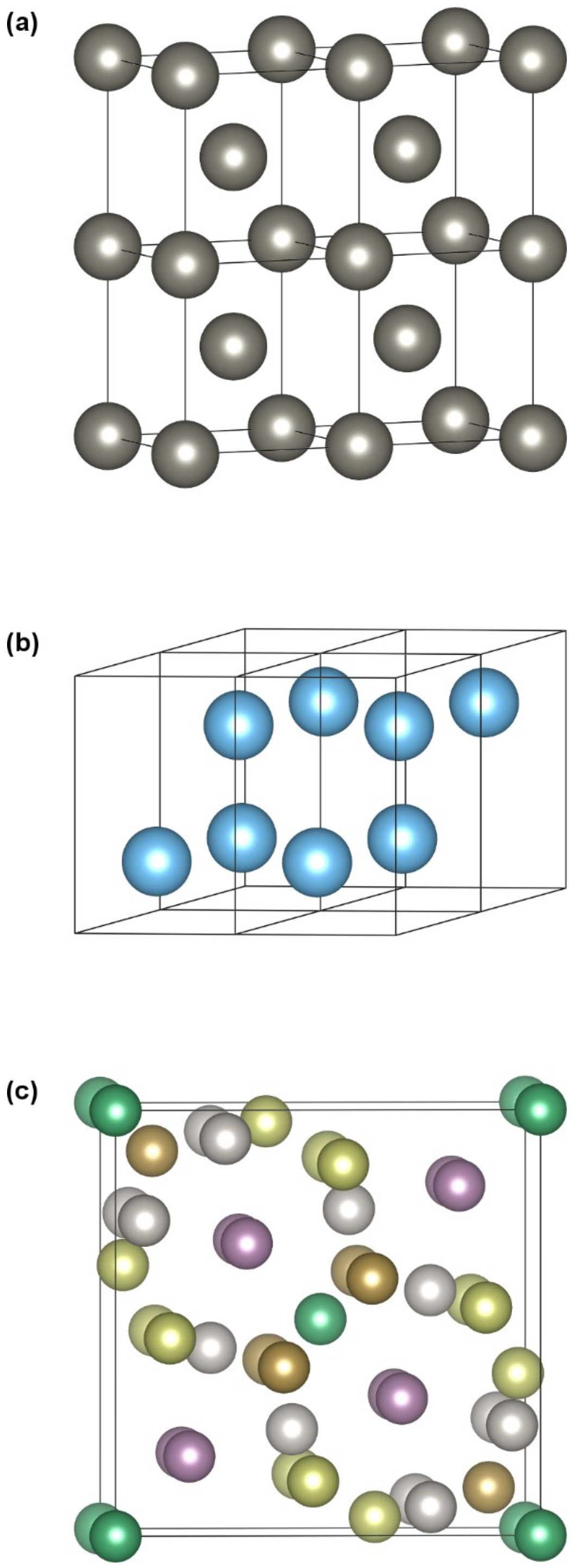

Figure 1: Superconducting high-entropy alloys adopt crystal structures such as (a) body-centred cubic, (b) hexagonal close-packed, and (c) $\sigma$ phase. Here, atom colours distinguish different crystallographic sites, which in high-entropy materials are each statistically occupied by some combination of the multiple constituent elements.

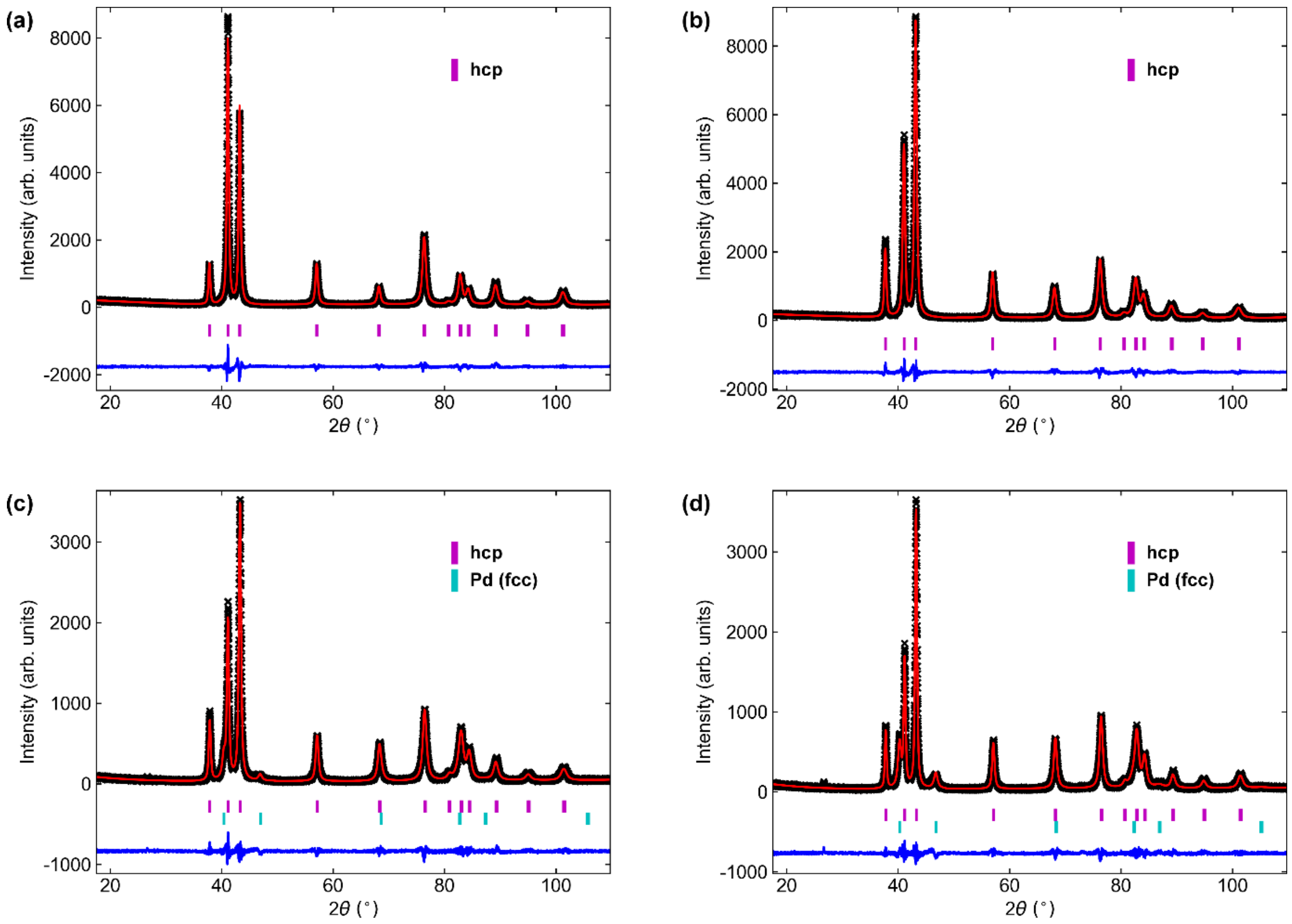

Figure 2: Le Bail fits to powder X-ray diffraction patterns of (a) MoReRuRhPt ($R_{wp}$ = 8.40%, $\chi^2$ = 1.83), (b) MoReRuIrPt ($R_{wp}$ = 8.09%, $\chi^2$ = 1.91), (c) MoReRuRhPd ($R_{wp}$ = 10.6%, $\chi^2$ = 1.50), and (d) MoReRuIrPd ($R_{wp}$ = 11.0%, $\chi^2$ = 1.60). The Pt-containing alloys crystallise as single-phase hcp HEAs, but as Pd is less soluble than Pt in a mixture of the remaining elements it is only partially incorporated into the HEA phase in the latter two compositions.

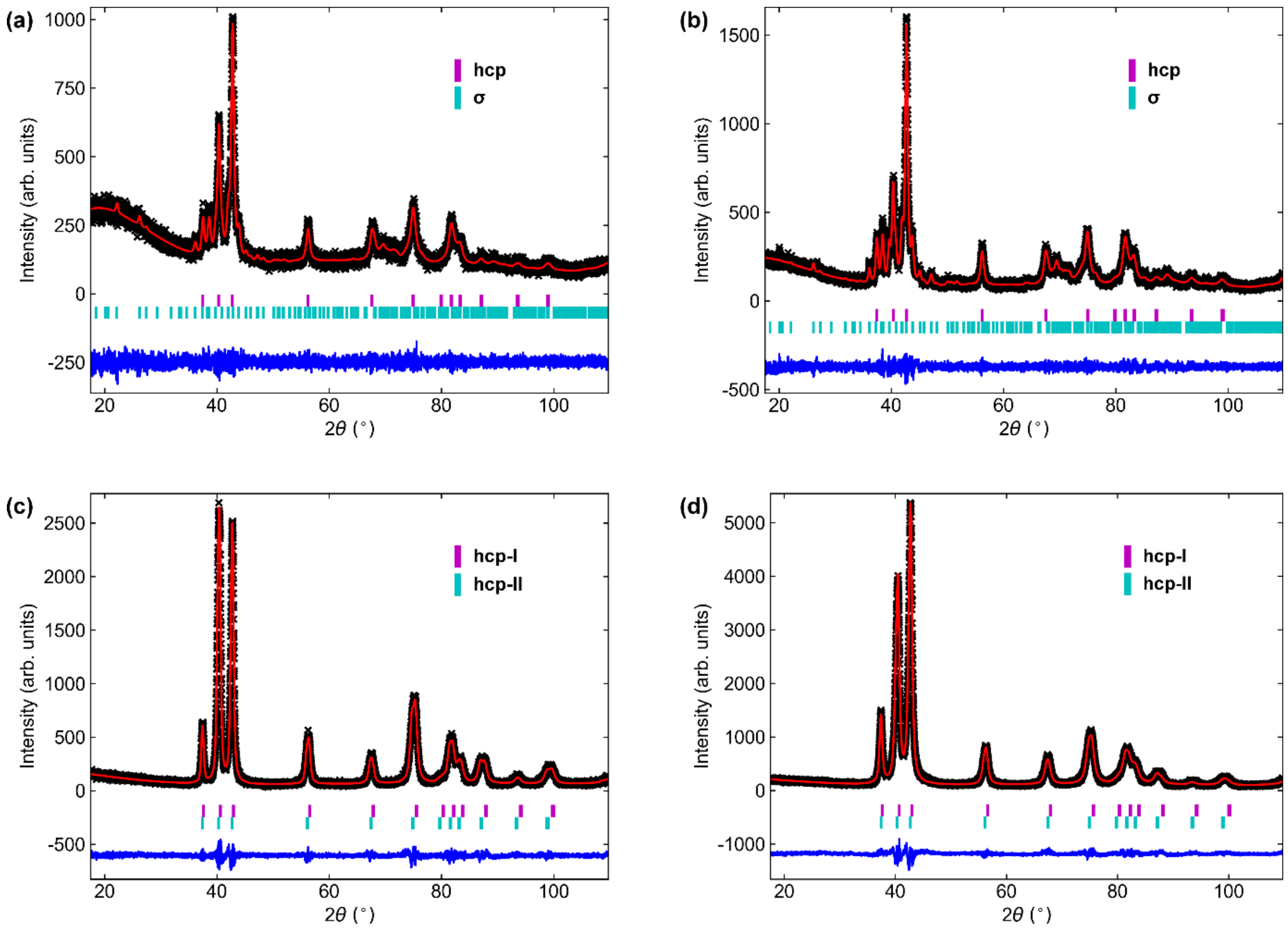

Figure 3: Le Bail fits to powder X-ray diffraction patterns of (a) NbMoReRuRh ($R_{wp}$ = 7.91%, $\chi^2$ = 1.06), (b) NbMoReRuIr ($R_{wp}$ = 8.78%, $\chi^2$ = 1.23), (c) NbMoReRuRhPt ($R_{wp}$ = 9.05%, $\chi^2$ = 1.37), and (d) NbMoReRuIrPt ($R_{wp}$ = 8.31%, $\chi^2$ = 2.11). These compositions give larger values of $\delta$ and $\Delta\chi$ than those in Figure 2 (Table 1), and in the first two here this manifests in the crystallisation of mixed hcp and $\sigma$ phases. In the latter two here the addition of a sixth element counters those enthalpic effects by increasing $\Delta S_{mix}$, but the temperature-dependence of entropic stabilisation leaves it insufficient to prevent phase separation on cooling.

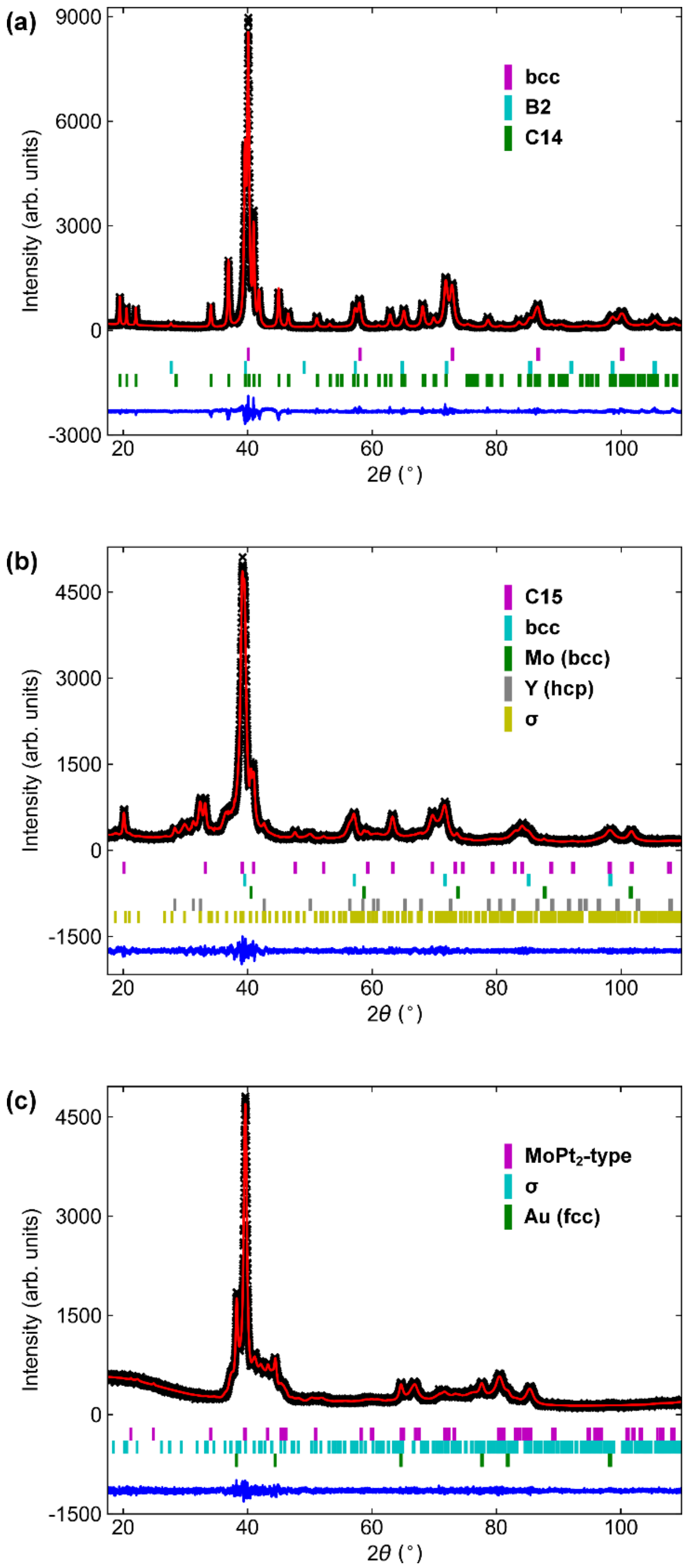

Figure 4: Le Bail fits to powder X-ray diffraction patterns of (a) ZrNbMoReRu ($R_{wp}$ = 9.81%, $\chi^2$ = 2.87), (b) YZrNbMoRe ($R_{wp}$ = 6.11%, $\chi^2$ = 1.35), and (c) NbMoRuPtAu ($R_{wp}$ = 5.72%, $\chi^2$ = 1.14). Large values $\delta$ and $\Delta\chi$ (Table 1) prevent these systems from crystallising as single-phase HEAs.

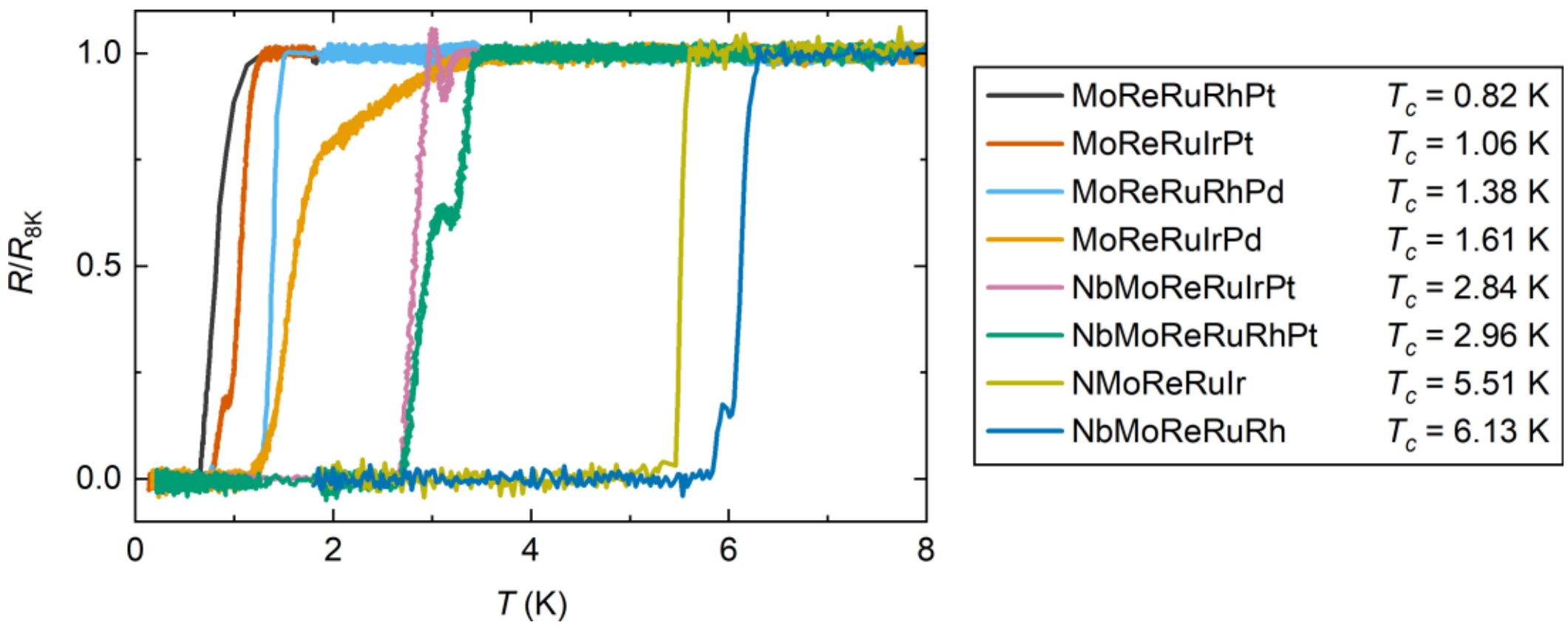

Figure 5: Electrical resistance of the new hcp and hcp-$\sigma$ HEAs we report. To facilitate comparison the measured resistance $R$ of each material has been normalised against its metallic-state value at 8 K. The superconducting critical temperature $T_c$ of each material is taken to be that at which $R/R_{8K} = 0.5$. Noise in the measurement of MoReRuRhPd was reduced by taking a weighted 10-point average for each datum.

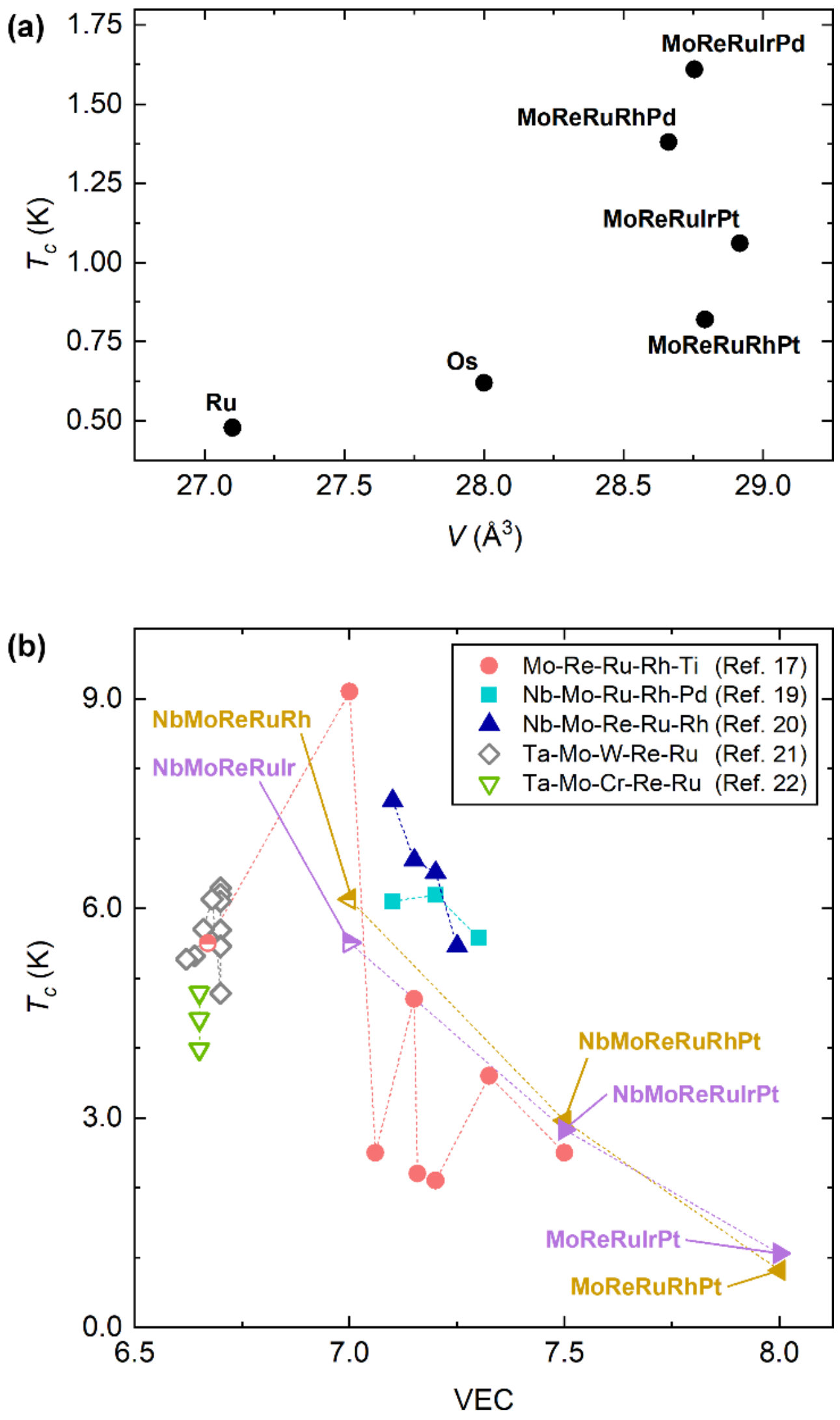

Figure 6: (a) The critical temperature of superconductivity $T_c$ in the hcp, VEC = 8 elements (unit cell volumes taken from [36], $T_c$ values from [44]) and our comparable HEAs. $T_c$ is dependent on the volume of the unit cell, $V$, through its effect on the density of states at the Fermi level. It is also affected by the band-filling, as described by Matthias, and the anomalously high critical temperatures of the two Pd-containing HEAs are the result of the hcp phases not fully dissolving the equimolar amount of Pd, meaning that their VECs are actually less than 8. (b) Comparison of the critical temperatures of different hcp and $\sigma$ HEAs. These show a general Matthias-like dependence on VEC, supplemented by compositional effects. Filled symbols indicate a hcp single-phase material, open symbols indicate a $\sigma$ single-phase material, and half-filled symbols indicate a mixed hcp- and $\sigma$-phase material. Dashed lines are guides for the eye.